\documentclass[twocolumn,superscriptaddress,showpacs,floatfix]{revtex4-1}
\usepackage{graphicx}
\usepackage{amssymb,amsmath,amsbsy,amsfonts,bm}
\usepackage{epsfig}
\usepackage{multirow}
\usepackage{tikz}
\usepackage{hyperref}
\usepackage{gensymb}
\usepackage{threeparttable}

\usepackage{graphicx}
\usepackage[caption=false]{subfig}
\usepackage{subfig}
\usepackage{lipsum}
\usepackage{float}
\usepackage[normalem]{ulem}%

\usepackage[percent]{overpic}
\usepackage{color}

\definecolor{darklavender}{rgb}{0.45, 0.31, 0.59}
\definecolor{amethyst}{rgb}{0.6, 0.4, 0.8}

\newcommand{\vsp}{v_0}

\newcommand{\perf}{\left < v_y \right >/v_0}

\newcommand{\pos}{{\mathbf{r}}(t)}
\newcommand{\ori}{\hat{\mathbf{u}}(t)}
\newcommand{\mot}{\mu(\mathbf{r})}
\newcommand{\muav}{\left < \mu \right >}

\DeclareMathOperator*{\argmax}{\arg \max}

\begin{document}
\title{Active particles using reinforcement learning to navigate in complex motility landscapes}

\author{Paul A. Monderkamp}
\email{paul.monderkamp@hhu.de}
\affiliation{Institut f\"ur Theoretische Physik II: Weiche Materie, Heinrich-Heine-Universit\"at D\"usseldorf, Universit\"atsstra{\ss}e 1, 
40225 D\"usseldorf, Germany}
\author{Fabian Jan Schwarzendahl}
\affiliation{Institut f\"ur Theoretische Physik II: Weiche Materie, Heinrich-Heine-Universit\"at D\"usseldorf, Universit\"atsstra{\ss}e 1, 
40225 D\"usseldorf, Germany}
\author{Michael A. Klatt}
\affiliation{Institut f\"ur Theoretische Physik II: Weiche Materie, Heinrich-Heine-Universit\"at D\"usseldorf, Universit\"atsstra{\ss}e 1, 
40225 D\"usseldorf, Germany}
\author{Hartmut L\"owen}
\affiliation{Institut f\"ur Theoretische Physik II: Weiche Materie, Heinrich-Heine-Universit\"at D\"usseldorf, Universit\"atsstra{\ss}e 1, 
40225 D\"usseldorf, Germany}
\begin{abstract}
As the length scales of the smallest technology 
continue to advance beyond the micron scale 
it becomes increasingly important to equip robotic components with the means for intelligent and autonomous decision making with limited information. With the help of a tabular Q-learning algorithm, we design a model for training a microswimmer, to navigate quickly through an environment given by various different scalar motility fields, while receiving a limited amount of local information. We compare the performances of the microswimmer, defined via time of first passage to a target, with performances of suitable reference cases. 
We show that the strategy obtained with our reinforcement learning model indeed represents an efficient navigation strategy, that outperforms the reference cases. By confronting the swimmer with a variety of unfamiliar environments after the finalised training, we show that the obtained strategy generalises 
to different classes of random fields.

\end{abstract} 

\maketitle

\section{introduction}

Technological advances in producing micron sized swimmers and robots
give hope for applications to minimal invasive medicine~\cite{nelson2010microrobots}. 
The possibilities reach from targeted drug delivery, over  
material removal in minimal invasive surgery, to telemetric 
applications where microrobots transmit information that is otherwise 
hard to obtain. In all of these examples, microrobots need to find 
a specific target, e.g., the location to which a drug needs to be delivered,
or an infected piece of tissue that needs to be surgically extracted. 
In order to find these targets, usually only local information about the 
surrounding environment of the robot is given. The robots might need 
to travel through a complex network of veines or pass through mucus, 
which makes navigation challenging. 
Hence, smart navigation strategies for microswimmers need to be found. 
Here, we develop intelligent strategies, that utilise only limited local information, for microswimmers in a complex motility field by
employing reinforcement machine learning techniques.

Recently, 
 machine learning techniques have been applied to active and soft matter systems~\cite{cichos2020machine,clegg2021characterising,falk2021learning}. 
Specifically, artificial microswimmers, which have been studied intensely~\cite{bechinger2016active},
might be used for technological applications such as decontamination of polluted water \cite{gao2014environmental}, 
or minimal invasive surgery \cite{abdelmohsen2014micro,patra2013intelligent,nelson2010microrobots,you2018intelligent}.
Active particles have been taught to navigate in different environments, for example 
optimal paths in force fields~\cite{schneider2019optimal,yang2018optimal,la2014multirobot} 
or flow~\cite{liebchen2019optimal,daddi2021hydrodynamics,zanovello2021target,nasiri2022reinforcement} have been computed.
Related to the latter, gliders have learned to navigate in a turbulent flow \cite{reddy2016learning,reddy2018glider} and
microswimmers learned a complex flow field~\cite{colabrese2017flow,colabrese2018smart,gustavsson2017finding,alageshan2020machine,qiu2020swimming,biferale2019zermelo}.
In experiments, reinforcement learning has been applied to microswimmers~\cite{muinos2021reinforcement} and artificial visual perception has been given to active colloids~\cite{lavergne2019group}.\\

The motility of active particles is strongly influenced by the surrounding medium~\cite{breoni2020active}, and in particular, viscous landscapes have been studied~\cite{datt2019active,liebchen2018viscotaxis}, giving rise to viscotaxis. Furthermore, active particles can be steered with an
orientation dependent motility~\cite{sprenger2020active}.

\begin{figure}[ht]
\begin{center}
\includegraphics[width=\linewidth]{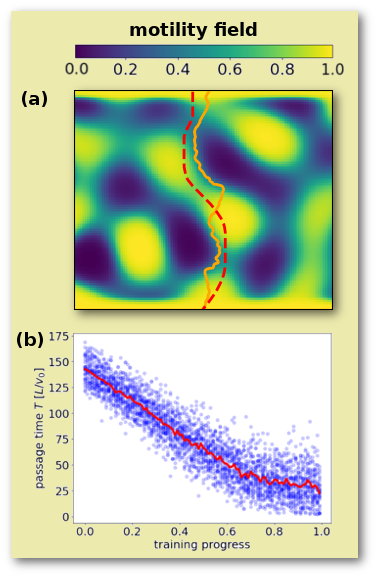}
\caption{
Training an active Brownian particle to navigate, with limited local information, through a complex random environment given by a scalar motility field. 
\textbf{(a)}: Typical trajectory after the finalised training procedure (solid orange line) and optimal trajectory obtained with the help of \textit{Dijkstras (shortest path) algorithm} (dashed red line). \textbf{(b)}: Average passage time through the simulation box (from bottom to top) at each percent of the training procedure (red line), sampled from $30$ independent training procedures. Each blue dot denotes an average over $10^4$ trajectories. }
\label{fig_concept}
\end{center}
\end{figure}

In this paper, we teach an active particle that has only local information about its environment to navigate through a complex motility field. A reinforcement learning technique (Q-learning, see Sec.~\ref{sec_qalg}) that requires a limited amount of data storage is used, making it usable for real life applications. Over the training time, the Q-learning active Brownian particle (QABP) learns to solve different realisations of a random environment, with increasing success (Fig.~\ref{fig_concept}.\textbf{(b)}).
At the end of training, the particle outperforms a simple active Brownian particle (ABP), and comes close to the globally optimal path (Fig.~\ref{fig_concept}.\textbf{(a)}) with only local information. Furthermore, once the particle has learned a strategy, we place it in qualitatively different environments, in which it still finds an almost optimal path. 

\section{methods}

\subsection{Equations of motion}

We model the swimmer as an overdamped active Brownian particle in two dimensions with position $\pos$ and orientation $\ori = (\cos \phi(t), \sin \phi(t))$. It exerts a space-dependent self-propulsion velocity $\vsp \mot$ along its orientation. $\mot \in (0,1]$ represents the motility field around the particle, such that the particle velocity is bound between $0$ and the self-propulsion velocity $v_0$. In order to perform intelligent navigation, the Q-learning active Brownian particle (QABP) is capable of optionally rotating itself with an angular velocity $\omega_Q(\pos,t) = \pm\omega_0$ in either direction. Accordingly, the equations of motion are

\begin{equation}
    \label{eq_rdot}
    \dot{\mathbf{r}}(t) = \vsp \mu(\bm{r}(t)) \ori, 
\end{equation}
\begin{equation}
    \label{eq_phidot}
    \dot{\phi}(t) = \omega_Q(\bm{r}(t),t) + \sqrt{2D_r} \xi,
\end{equation}
where $\xi$ represents Gaussian white noise exerted from the solvent environment on the orientation of the particle, with $\left < \xi(t)\right > = 0$ and $\left < \xi(t) \xi(t') \right > = \delta(t-t')$.

The swimmer moves within a box of side lengths $L_x=100 v_0 \tau_Q$ and $L_y=85 v_0 \tau_Q$, where $\tau_Q$ denotes the characteristic time scale of an intelligent decision (see Sec. \ref{sec_qalg}). The reinforcement learning problem is defined by navigating as quickly as possible from the bottom hard wall to the top hard wall of the box. 
Depending on the specific model of $\mot$, either reflecting or periodic boundary conditions in horizontal direction are used.
The QABP and the reference case of the ABP start their trajectories at $\mathbf{r}(t=0) = (0.5L_x,0)$, oriented upwards ($\hat{\mathbf{u}}(t=0) = (0,1)$).
The swimmers are trained in a motility field that is generated with a Gaussian random wave model (GRW), with wave vectors $\bm{k}_n$ of wavelength $\left\| \mathbf{k}_n \right\| = 20/L_y$ (for details see Supplemental Material Sec. 1 \cite{SI}).
The GRW gives isotropic non-periodic random waves, a typical example is shown Fig.~\ref{fig_concept}(a).
A remarkable property of this model is an optimal suppression of density fluctuations above a
certain wavelength, a property known as stealthy
hyperuniformity~\cite{torquato_hyperuniform_2018, ma2017random, chen2021hyperuniform,  klatt2022characterization}. 
The use of stealthy hyperuniform models as randomly generated motility fields has
the advantage that the formation of large clusters of low motility zones is suppressed.
This typically results in the presence of several paths through the environment
along which the passage time is comparable, which greatly facilitates the
learning of the QABP.

\subsection{Q-learning algorithm}
\label{sec_qalg}

\begin{figure}[t]
\begin{center}
\includegraphics[width=1.0\linewidth]{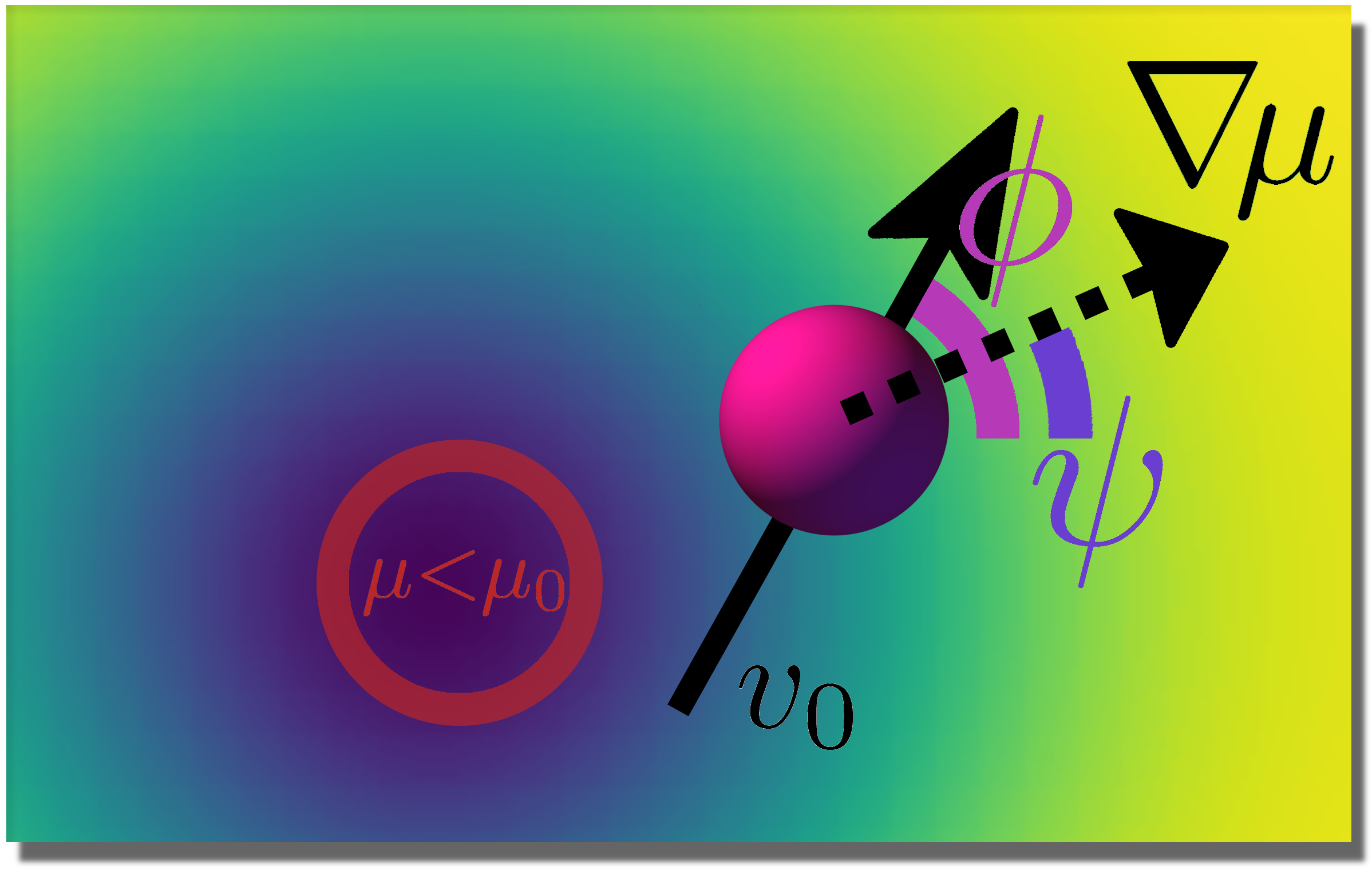}
\caption{
Schematic of the local information, that the QABP receives for the decision making.
The swimmer knows the discretised polar angle of its own orientation $\hat{\mathbf{u}} = (\cos(\phi),\sin(\phi))$, and the polar angle $\psi$ of the local gradient $\nabla \mu$ of the motility field. Further, it has information about low motility zones, defined by $\mu < \mu_0$, as indicated by the red circle. 
}

\label{fig_modelfig}
\end{center}
\end{figure}

To enable swimmer navigation within the simulated physical environment, a tabular Q-learning algorithm~\cite{sutton2018reinforcement} is superimposed on the Brownian dynamics simulation, giving the QABP the ability for self rotation through the torque expressed by $\omega_Q$. Such an algorithm is characterised by a matrix table $\mathcal{Q}$, which encompasses the strategy and learned experience by the active agent. This matrix functions as a decision matrix, where the rows represent all possible discrete states, in which the swimmer can reside and the columns represent all possible discrete actions. At any time $t$ of action, the swimmer checks its current state $i$, and performs the action $A_i$ corresponding to the highest value within row $i$ of the matrix: 
\begin{equation}
    A_i = \argmax_j \mathcal{Q}_{ij}(t).
    \label{eq_action}
\end{equation}
In our model, the swimmer has information (see Fig.~\ref{fig_modelfig}) about its own orientation $\phi(t)$ and the polar angle $\psi (\pos)$ of the local gradient of the motility field $\nabla\mu(\pos)$. Furthermore, it knows whether $\mu(\pos)$ is above or below a threshold value $\mu_0 = 0.25$ (see Supplemental Material Sec. 3 \cite{SI}). 

The orientational dynamics of the QABP (see Eq. \ref{eq_phidot})
are approximated by run and tumble dynamics, where the swimmer tumbles in each integration time step $t_n = n\Delta t$ with a probability $P_{\mathrm{tumble}}=2D_r\Delta t/(\Delta \phi)^2$, where $\Delta \phi = 2\pi/M_\phi$. 
The local gradient direction and swimmer orientation are discretised on the unit circle with $M_\phi = M_\psi = 12$. All possible combinations of discrete orientations and gradient directions as well as the binary information about the velocity form the complete state space of the Q-learning algorithm. With a given periodicity $\tau_Q = 10\Delta t$, the swimmer takes action, by rotating itself in either direction by $\Delta\phi$, or not rotating, depending on the decision matrix $\mathcal{Q}$.
This rotation defines an effective angular velocity $\omega_0 = \Delta\phi/\tau_Q = 2\pi/(N_\phi \tau_Q)$.

In order to obtain a decision matrix, which represents a good strategy for navigating through the complex environment given by $\mot$, $\mathcal{Q}$ is optimised over the course of $10^6$ episodes, i.e., trajectories. Before the training procedure, $\mathcal{Q}$ is initialised with zero values.  
For each episode in the learning phase, the trajectory of the swimmer is simulated until it either reaches the top of the box, swims into a 
region with $\mu < 0.5\mu_0 = 0.125$ or the travel time surpasses an upper bound $T_{\mathrm{max}}$, obtained by 100 times the time of a comparable optimal trajectory, obtained with 
\textit{Dijkstra's algorithm} (see Supplemental Material Sec. 2 \cite{SI}). 
To obtain a navigation strategy, as general as possible,
$10^3$ realisations of the random environment are used over $10^6$ episodes.
The results are sampled, with the trained QABP, on $10^3$ new environments.\\

During training, when an action $j$ is performed, the QABP transitions from state $i$ to $i'$.  
Then $\mathcal{Q}$ is updated, 
following the update formula
\begin{equation}
    \mathcal{Q}^{new}_{ij} = \mathcal{Q}_{ij} + \alpha \left (R + \gamma \max_k (Q_{i'k}) - \mathcal{Q}_{ij} \right ).
\end{equation}
Here, $\alpha,\gamma \in [0,1]$  denote hyperparameters of the learning algorithm and $R$ denotes the sum of the specific numeric rewards, that the active agent obtained through performing the current action $j$. 
The learning rate 
$\alpha$ is initialised at $10^{-4}$ and linearly decreases to $10^{-5}$ at the end of the training, reinforcing the reliability of $\mathcal{Q}$ with proceeding learning. 
The term $\gamma \max_j (Q_{i'j})$ incorporates the highest entry in a row from the following state into the 
current $\mathcal{Q}_{ij}$, estimating the future reward. Since a reasonably different behaviour in neighbouring states is expected across the swimmers state space,  $\gamma=0.3$ is used.
During training an $\epsilon$-greedy policy is used. Here, random actions are chosen with probability $\epsilon=1$ at the beginning of training, then during training $\epsilon$ is decreased linearly to $0$ such that Eq.~\eqref{eq_action} is used for any decision at the end of training (see Supplemental Material Sec. 5 \cite{SI}).
\\ 

In order to navigate efficiently the swimmer is rewarded once it reaches the top of the simulation box. Further, it is punished when it enters a low motility region, or if its displacement is very small (for reward details Supplemental Material Sec. 4. \cite{SI}).

\section{Results}

\begin{figure*}[htp]
\begin{center}
\includegraphics[width=1.0\linewidth]{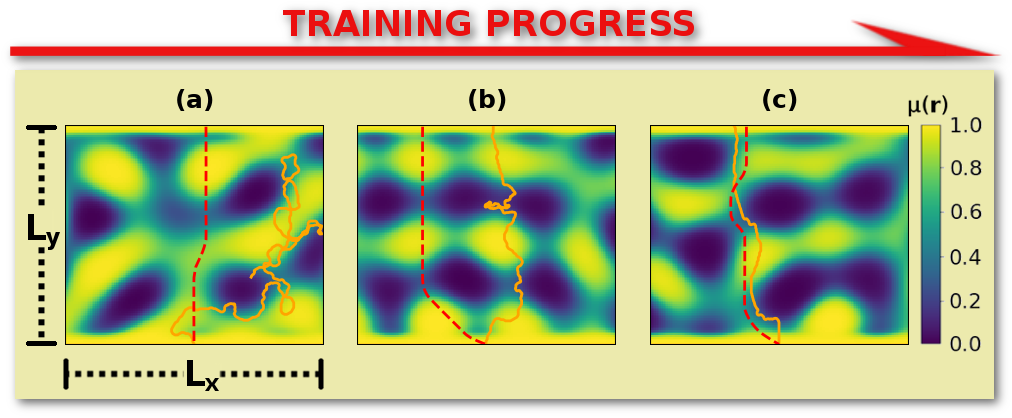}
\caption{Three typical trajectories (orange solid lines) in three motility fields $\mot$ over the course of the training procedure of the Q-learning active Brownian particle (QABP). 
The particle's objective is to cross the box from bottom to top as fast as possible.
The environments are generated with the help of modified isotropic Gaussian random waves. The probablities to perform random actions are  \textbf{(a)}: $\epsilon = 1.0$, \textbf{(b)}: $\epsilon \approx 0.58$, \textbf{(c)}: $\epsilon \approx 0$. The red dashed line highlight the optimal trajectories obtained with \textit{Dijkstras algorithm} (see Supplemental Material Sec. 2 \cite{SI}). 
Motility fields are generated as modified isotropic Gaussian random waves and in the horizontal direction reflecting boundary conditions are employed.}
\label{fig_trajfig}
\end{center}
\end{figure*}

To give an intuition on the development of the strategy, three characteristic trajectories from different stages of the learning process are shown in Fig.~\ref{fig_trajfig}. For visual reference, each panel additionally shows a globally optimal trajectory obtained via \textit{Dijkstra's algorithm}. In Fig.~\ref{fig_trajfig}.\textbf{(a)} we show a trajectory from an episode early in the training procedure. Since the Q-learning active Brownian particle (QABP) has yet to learn about its environment, the probability $\epsilon$ to perform a random rotation in either direction 
is close to 1. Accordingly, the trajectory is similar to that of a common active Brownian particle (ABP).
Due to the indecisiveness of the QABP at this explorative stage, the episode terminates eventually by entering a low motility zone.\\

Figure \ref{fig_trajfig}.\textbf{(b)} depicts a trajectory from an episode 
halfway through the training procedure. The corresponding probability to perform random actions $\epsilon$ is approximately $0.45$.
The trajectory shows randomness, through rotational diffusion as well as random active rotation. Despite the fact that more than half of the actions are randomly chosen, it is visible, how the QABP displays noticeable competence of avoiding the regions with $\mu \ll 1$ to reach the finish line.
Finally, the trajectory in Fig.~\ref{fig_trajfig}.\textbf{(c)} shows the dynamics of the QABP after the learning procedure when $\epsilon =0$. The QABP swims decisively in vertical direction, such that the trajectory exhibits little dents, thereby 
maneuvering around the low motility zone in its path. 

\subsection{Quantitative performance}
\label{sec_performance}

\begin{figure}[t]
\begin{center}
\includegraphics[width=\linewidth]{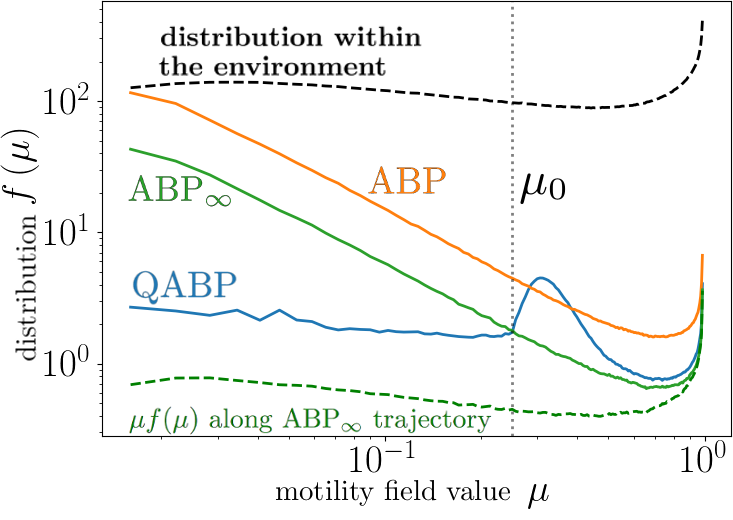}
\caption{Distributions of scalar values $\mu(\mathbf{r}(t))=\left \| \dot{\mathbf{r}}\right \|/v_0$ along trajectories with $\muav \approx 0.557$ and $\mathcal{P}_{\mathrm{rot}} \approx 46$, sampled from 15 simulations with 1000 trajectories each. The lines indicate ABP (solid orange), straight trajectory, i.e. ABP$_\infty$ (solid green), and intelligent swimmer (QABP, solid blue). The dashed black line shows the distribution of the $\mot$-values within the environment, averaged from $1000$ independent $\mot$. The dashed green line corresponds to $\mu f(\mu)$ of the ABP$_\infty$ trajectory. The dotted vertical line denotes the velocity state threshold $\mu_0 = 0.25$.}
\label{fig_velocities_hist}
\end{center}
\end{figure}

Optimising the navigation through the environment, given by the motility field $\mot$, relies on the balancing between two opposing principles: minimising the length of the path 
while simultaneously 
maximising the instantaneous velocity $v_0\mot$. Formally, this problem is solved by the solution that minimises the following functional 
\begin{equation}
\label{eq_Tfunc}
    T[\mathbf{c}] = \int_{\mathbf{c}_0}^{\mathbf{c}_1}    \frac{\left\| \bm{\dot c}(t)\right\|}{v_0\mu(\mathbf{c}(t))} \mathrm{d} t ,
\end{equation}
which is the passage time $T$ from the starting point $\mathbf{c}_0$ to any point on the finish line $\mathbf{c}_1$. Here, $\bm{c}(t)$ is a curve through the environment, parameterised by $t$.
A convenient figure of merit for the performance of the swimmer along any trajectory $\pos$ is defined as
\begin{equation}
    \label{eq_performance}
   \frac{v_y}{v_0} := \frac{L_y}{T[\pos]v_0}.
\end{equation}

This quantity is in the interval $(0,1]$ for any trajectory. After each training procedure, the resulting $\mathcal{Q}$ is tested in $10^3$ trajectories on an independent realisation $\mot$ respectively. The general performance of the resulting strategy, encoded in 
$\mathcal{Q}$, is determined by averaging over the $10^3$ trajectories, giving $\left < v_y \right >/v_0$. The performance will depend on the average motility $\langle\mu\rangle$ and the rotational P\'eclet number, which is defined as 
\begin{equation}
    \mathcal{P}_{\mathrm{rot}}  = {(\Delta\phi)^2}/{2D_{\mathrm{rot}}\tau_Q}
\end{equation} 
comparing the typical time scale of the rotational diffusion to that of the intelligent active rotation.\\

The total passage time $T$ (see Eq. (\ref{eq_Tfunc})) can be calculated from the non-normalised distribution of velocities $f(\mu)$, sampled with a time step 
of $\tau_Q$, along a given trajectory as
\begin{equation}
\label{eq_TfromN}
    T = \tau_Q \int_0^1 f(\mu) d\mu.
\end{equation}
To quantify the behaviour of the individual swimmers, the frequencies $f(\mu)$ of the scalar values of $\mot$ are shown 
in Fig.~\ref{fig_velocities_hist}. The data is averaged from $15$ independent simulation runs with $10^3$ trajectories each.
Additionally $\mu f(\mu)$ is shown for the ABP, for a straight line (i.e., an ABP in the limit of vanishing diffusion denoted by ABP$_\infty$), and for the distribution of function values $\mot$ within the environment, averaged from $10^3$ independent motility fields.
More specifically, the distribution of the ABP$_\infty$ trajectory (green solid line) counts the frequencies of function values of $\mot$ along a vertical line, sampled in time. Due to the longer retention time at slower velocities, we obtain the distribution in space, sampled at constant distances, as 
$\mu f(\mu)$. Since the data is averaged over $1.5\times 10^4$ different environments, the corresponding $\mu f(\mu)$ (green dashed line) is proportional to the distribution of the values in the total field (black dashed line). On the other hand, the ABP$_\infty$ reflects the limit of the ABP, for vanishing diffusion. As visible in the plot, the respective curves are proportional to each other, emphasising that the performances of the ABP is tangible through consideration of the motility field alone. Observing the QABP case, it is visible that the distribution in the lower velocities is approximately constant and lies orders of magnitude below both ABP cases.
Instead, $f(\mu)$ displays a significant peak above the velocity state threshold indicated by the vertical line at $\mu_0 = 0.25$. This exemplifies how the navigation strategy of the QABP relies on circumvention of the low motility zones, through higher motility regions, thereby elongating the trajectory, but saving time through the faster swimming. \\

\begin{figure}[t]
\begin{center}

\subfloat[$\muav \approx 0.557$]{
\includegraphics[width=\linewidth]{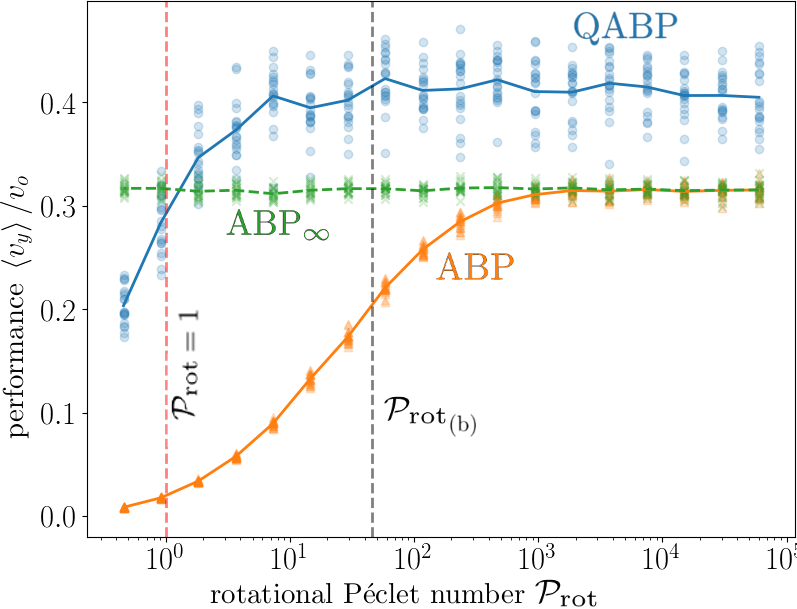}
}

\subfloat[$\mathcal{P}_{\mathrm{rot}} \approx 46$ ]{
\includegraphics[width=\linewidth]{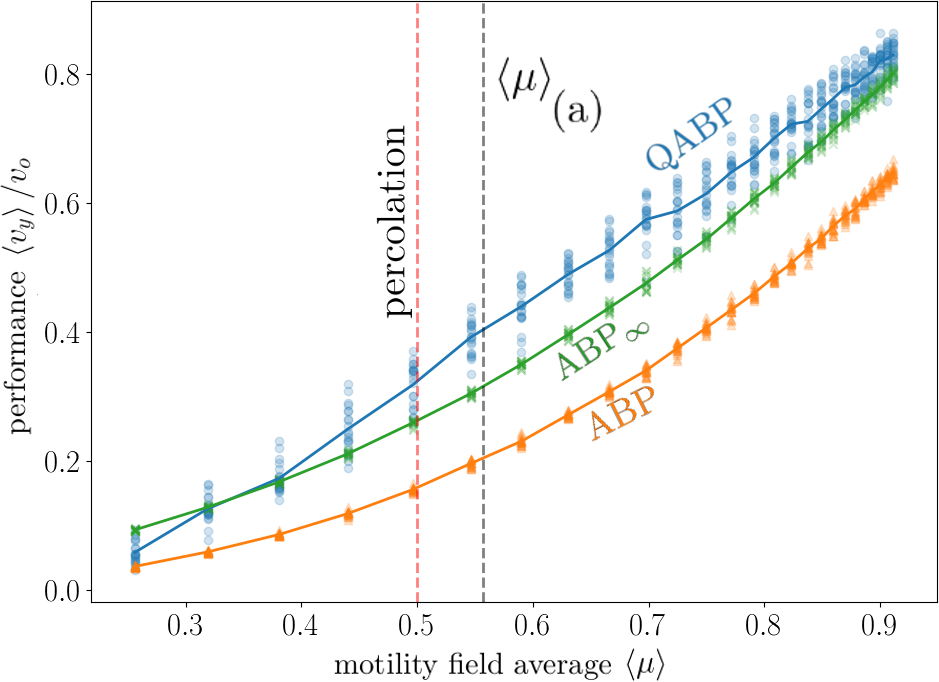}}
\caption{Performances $\left < v_y \right >/v_0$ of the QABP as well as an active Brownian particle (ABP) and a particle swimming in a straight line without diffusion (ABP$_{\infty}$) as reference swimmers. The scattered data data points indicate a simulation with $10^3$ trajectories each. The solid lines denote the averages over the respective ten data points per parameter set. (a) shows the performance for varying rotational P\'eclet number $\mathcal{P}_{\mathrm{rot}}$ and (b) for varying environment $\langle\mu\rangle$.}
\label{fig_performances}
\end{center}
\end{figure}

The performances $\perf$ of the QABP, and the two reference cases (ABP, ABP$_\infty$) are shown as a function of $\mathcal{P}_{\mathrm{rot}}$ in Fig.~\ref{fig_performances}.\textbf{(a)}, where $\muav \approx 0.557$ is chosen. 
For large $\mathcal{P}_{\mathrm{rot}}$, the $\perf$ of the ABP approaches the performance of the ABP$_\infty$. More specifically, the ABP is bounded by the ABP$_\infty$, and $\perf$ is monotonous in $\mathcal{P}_{\mathrm{rot}}$. For small $\mathcal{P}_{\mathrm{rot}}$, the performance of the ABP approaches $0$. For almost the whole parameter range the QABP is faster than both reference cases. Additionally $\perf$ seems to be independent from $\mathcal{P}_{\mathrm{rot}}$ for $\mathcal{P}_{\mathrm{rot}} \gtrsim 1$. This demonstrates the QABP's ability to steer against the kicks from rotational diffusion. Only for $\mathcal{P}_{\mathrm{rot}} \lesssim 1$, the QABP loses its ability to correct for rotational noise, and hence the performance declines with decreasing values of $\mathcal{P}_{\mathrm{rot}}$. 
Figure~\ref{fig_performances}~\textbf{(b)} shows the respective performances as a function the environment parameter $\muav$ for a constant $\mathcal{P}_{\mathrm{rot}} \approx 46$. For all three cases, $\perf$ increases with $\mu$. Once again, the QABP surpasses the ABP across all $\muav$, and the ABP$_\infty$ for most of the shown parameter range. For $\muav \approx 1$, the performances $\perf$ of both QABP and ABP$_\infty$ approach $1$, since the optimal trajectory becomes a straight line. For $\muav = 0.5$, the motility field $\mot$ displays a percolation transition of the low motility zones, and therefore the Q-learning model of avoiding low motility zones becomes less viable. It is visible in the figure, however, that the QABP surpasses the ABP$_\infty$ down to $\muav \approx 0.35$. 

\subsection{Generalisation to unfamiliar environments}
\label{sec_transfer}

In order to demonstrate the generality of the learned strategy, i.e. to show that the final result $\mathcal{Q}$ of the learning process transcends the specific implementation of our learning environment, we test the QABP's ability to navigate in other motility fields
that represent different types of long- and short-order order in space. More specifically, the swimmer is first trained on the previously used non-periodic Gaussian random fields with reflecting boundary conditions. After the learning procedure is finalised, the swimmer is placed in the respective unfamiliar environment.\\

A selection of the emerging trajectories ($\mathcal{P}_{\mathrm{rot}} \approx 46$) is shown in Fig.~\ref{fig_transferleistung}. All fields are generated such that $\muav \approx 0.557$. Additionally, a performance $\perf$ is given averaged from ten independent training procedures with $10^3$ trajectories each. For all fields, the QABP outperforms the references cases.

\subsubsection{Directional patterns}
The first instance in Fig.~\ref{fig_transferleistung}.\textbf{(a)} displays an environment 
that is periodic in horizontal direction. The horizontal components of the random wave vectors 
fulfill $\left | \mathbf{k}_n \cdot \hat{\mathbf{e}}_x \right | = 6\pi / L$. The absolute value of the vertical components $\left | \mathbf{k}_n \cdot \hat{\mathbf{e}}_y \right |$ are randomised uniformly in $[0,12\pi / L]$.
The QABP visually displays competence of maneuvering through the new environment. Furthermore, $\perf$ shows similar values as presented for the same parameters ($\mathcal{P}_{\mathrm{rot}} \approx 46$, $\muav \approx 0.557$) in Sec.~\ref{sec_performance}. This likely stems from the similarity of the wave vectors $\mathbf{k}_n$ and the resulting characteristic length scales in the environments $\mot$.

Figures~\ref{fig_transferleistung}~\textbf{(b)} and \textbf{(c)} show trajectories through periodic Gaussian random fields with $\mathbf{k}_n \cdot \hat{\mathbf{e}}_x  = 2\pi / L$ and $\mathbf{k}_n \cdot \hat{\mathbf{e}}_y  = \pm 8\pi / L$, respectively. It is visible through the global solutions, that the optimal paths are obtained by avoiding the low motility stripes through the periodicity of the box, persistently swimming in the same slanted direction. The performance of both environments with opposite parity are approximately equal. We thereby show, that the machine learning model is on average not subject to unexpected symmetry breaking. 

\subsubsection{Gray-Scott field}
Figure~\ref{fig_transferleistung}~\textbf{(d)} shows an environment obtained through integration of the Gray-Scott equations for reaction-diffusion \cite{mcgough2004pattern,gray1990chemical} (see Supplemental Material Sec. 1.A \cite{SI}). The typical length scales of the low motility zones obtained through this model are drastically smaller, than the previous examples. Furthermore, due to the entirely different algorithm, the functional form at the edges of the low motility zones is different. Despite these drastic differences to the original training data, the QABP displays the capability of efficiently maneuvering through the environment.

\subsubsection{Strongly clustering patterns}
All of the previously shown examples of scalar fields are based on hyperuniform models
with suppressed density fluctuations.
Next, we show that after training, the QABP also successfully navigates in random motility fields
with strong heterogeneities (based on non-hyperuniform models).

\begin{figure*}[t]
\begin{center}
\includegraphics[width=\textwidth]{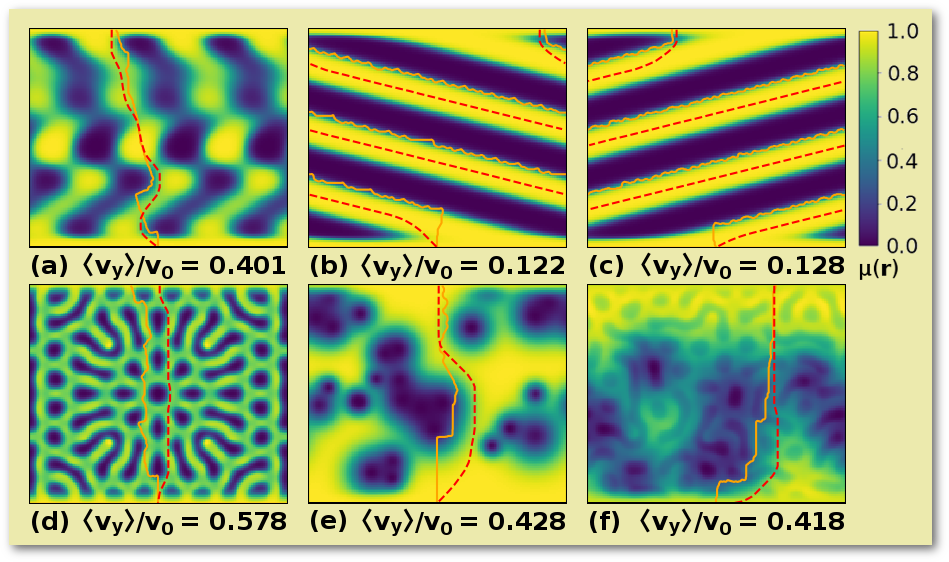}
\caption{Trajectories of QABPs, trained in non-periodic Gaussian random wave environments (see. Fig.~\ref{fig_trajfig}), placed in unknown environments, after the finalised learning process ($\mathcal{P}_{\mathrm{rot}} \approx 46$, $\muav \approx 0.557$). The globally optimised trajectories, generated with the help of \textit{Dijkstra's algorithm}, are shown as dashed red lines. The motility fields $\mot$ are generated with the help of \textbf{(a):} Gaussian random waves (see. Supplemental Material Sec. 1) periodic in horizontal direction, \textbf{(b,c):} Gaussian random waves periodic in both Carthesian directions,
\textbf{(d):} Gray-Scott model for reaction-diffusion
\textbf{(e):} $30$ Gauss peaks scattered randomly across the simulation box and 
\textbf{(f):} Eight Gauss peaks scattered randomly across the simulation box, superimposed with solutions to the randomly initialised Gray-Scott model.}
\label{fig_transferleistung}
\end{center}
\end{figure*}

Figure~\ref{fig_transferleistung}~\textbf{(e)} shows an environment, which is generated by scattering $30$ points $\mathbf{q}_k$ uniformly in the simulation box, and assigning each a characteristic length scale $\kappa$. Explicitly, the motility field is given by
\begin{equation}
    \label{eq_model0}
    \mot = \prod_ {k=1}^{30} \left [ 1 - \exp \left ( \frac{\left \| \mathbf{r} - \mathbf{q}_k \right \| ^2 }{\kappa^2}  \right ) \right ],
\end{equation}
with periodic boundary conditions.
The environment in Fig.~\ref{fig_transferleistung}.\textbf{(e)} displays several clusters of peaks. Nevertheless, we observe that the QABP finds a quick path through the environment by evading low motility zones of any size (for which it was efficiently trained in the stealthy hyperuniform Gaussian random wave model).

Finally, Fig.~\ref{fig_transferleistung}~\textbf{(f)}, shows a trajectory through a motility field, generated by scattering eight Gaussian peaks with amplitude $1$, combined by summation, with a randomly initialised solution to the Gray-Scott model with amplitude $0.1$ (see Supplemental Material Sec. 1.A \cite{SI}). This approach yields a motility field, with large low motility zones throughout the environment, overlaid with a more regular perturbation, that causes local gradients, which drastically influence local information. Even though, not having learned about the local structure, our results show, that the QABP noticeably interacts 
with the respective local gradients only at low swimming velocities. 
This result emphasises the significance of the inclusion of the swimming velocity in the QABPs state space.

\section{Conclusions}

In this work, we used a reinforcement learning algorithm to teach a microswimmer (QABP) to navigate through complex environments, given by scalar motility fields, 
that determine the local swimming velocity of the particle. 
Brownian dynamics simulations were used to investigate the dynamics of the QABP in this two-dimensional physical environment. 
To enable smart navigation, a tabular Q-learning algorithm was superimposed. The swimmer receives the ability to perform deterministic rotations, while only receiving local information about its environment.\\

First, modified Gaussian random waves were employed as motility fields. Two reference cases of an active Brownian particle (ABP) and a particle swimming in a straight line (ABP$_\infty$) were simulated and it was shown that the time of first passage of the ABP to a given target can be inferred from the motility field. The performance, i.e. the speed of finding the target, of the ABP, is bounded by the ABP$_\infty$, as the limit of low diffusion. We demonstrate, that our intelligent QABP outperforms both the ABP and ABP$_\infty$. To demonstrate the applicability of the resulting strategy, we test the ability of the QABP to solve different environments, generated with various algorithms, though only having learned the Gaussian random wave environment. The swimmers display competence of maneuvering through all the displayed examples of motility fields and again outperforms the ABP and ABP$_\infty$.
Our stealthy hyperuniform model provides a random yet relatively homogeneous environment
that is well suited for the initial training of the QABP.
Afterwards, the QABP also navigates without difficulty in non-hyperuniform fields
with strong heterogeneities.\\

Throughout this paper, we lay emphasis on the fact, that the final decision matrix only requires the microswimmer to know little local information about its environment. This will be of particular relevance to future microrobotic applications, where individual autonomous agents rarely possess the ability to capture information about the whole environment at once, and the amount of data storage if dictated by the size of the technical components. 
Future studies of this local algorithm can be extended to more complex problems such as the inclusion of hydrodynamic force fields \cite{liebchen2019optimal,nasiri2022reinforcement}.
The explicit inclusion of cargo uptake and delivery~\cite{ma2015catalytic,demirors2018active}, as well as the consumption of fuel, into the machine learning model, may be of interest to medical applications~\cite{nelson2010microrobots}. Swimming strategies, which combine a deterministic approach to the decision making, such as through our reinforcement learning, with undeterministic approaches, e.g., random actions (cf. Fig.~\ref{fig_trajfig}].\textbf{(b)}), may yield insight as models for biological microswimmers, that are motivated, for instance by the search for nutrients. 

\section*{acknowledgements}

The authors like to thank Jens Grauer for sharing his machine learning expertise and methodic advice. Without his impactful feeback, this paper would not exist in the current form. \\
This work was supported by the German Research Foundation (DFG) within project LO~418/25-1.

\section*{Data Access Statement}

Research data supporting this publication will be provided upon reasonable request via e-mail to the corresponding author.

\section*{Conflict of Interest declaration}
The authors declare, that there are no conflicts of interest to disclose.

\bibliography{bibliog}\end{document}


\title{Supplemental Material: Active particles using reinforcement learning to navigate in complex motility landscapes}
\date{\today}
\def\hhu{\affiliation{Institut f\"ur Theoretische Physik II: Weiche Materie, Heinrich-Heine Universit\"at D\"usseldorf, Universit\"atsstr.~1, 40225 D\"usseldorf, Germany}}
\def\inn{\affiliation{Institut f\"ur Theoretische Physik,
  Universit\"at Innsbruck, Austria}}
\def\tud{\affiliation{Theorie Weicher Materie, Fachbereich Physik, Technische Universität Darmstadt, Hochschulstraße 12, 64289 Darmstadt, Germany}}

\author{Paul A. Monderkamp}
\email{paul.monderkamp@hhu.de}
\affiliation{Institut f\"ur Theoretische Physik II: Weiche Materie, Heinrich-Heine-Universit\"at D\"usseldorf, Universit\"atsstra{\ss}e 1, 
40225 D\"usseldorf, Germany}
\author{Fabian Jan Schwarzendahl}
\affiliation{Institut f\"ur Theoretische Physik II: Weiche Materie, Heinrich-Heine-Universit\"at D\"usseldorf, Universit\"atsstra{\ss}e 1, 
40225 D\"usseldorf, Germany}
\author{Michael A. Klatt}
\affiliation{Institut f\"ur Theoretische Physik II: Weiche Materie, Heinrich-Heine-Universit\"at D\"usseldorf, Universit\"atsstra{\ss}e 1, 
40225 D\"usseldorf, Germany}
\author{Hartmut L\"owen}
\affiliation{Institut f\"ur Theoretische Physik II: Weiche Materie, Heinrich-Heine-Universit\"at D\"usseldorf, Universit\"atsstra{\ss}e 1, 
40225 D\"usseldorf, Germany}

\maketitle

\section{Learning environments}
\label{sec_env}
We train the swimmer in a Gaussian random wave model (GRW). To construct an appropriate motility field, we use a sinoidal function 
\begin{equation}
\label{eq_grw}
    g(\mathbf{r}) = \sum_{n=1}^{M} \cos \left ({\mathbf{k}_n\mathbf{r}} + \eta_n\right ),
\end{equation}
where the $\mathbf{k}_n$ and random wave vectors with $\left\| \mathbf{k}_n \right\| = 20/L$, drawn from an isotropic probability distribution and the $\eta_n$ are random phases, uniformly distributed in $[0,2\pi)$. $M$ denotes the number of superposed harmonic waves. We pick $M=100$, which is sufficient to obtain peaks with the shape of Gaussian bell curves. The motility field $\mot_{GRW}$ is obtained from $g(\mathbf{r})$ via

\begin{equation}
\label{eq_grw_mod}
    \mot_{GRW} = \frac{1}{1+ \exp\left ( \lambda \left [ 0.5 - \left [ \tilde{g}(\mathbf{r}) \right ]^\nu \right ] \right )},
\end{equation}
where $\tilde{g}(\mathbf{r})$ is obtained from $g(\mathbf{r})$ via renormalising to the interval $[0,1]$. 
We vary the exponent $\nu$, in order to shift the whole function towards $0$ or $1$, and therefore control the average function value $\left < \mu \right >$, 
by which the difficulty of a given field is characterised. Here, $\left < \mu \right >$ denotes the average over space and independent realisations of the field $\mot$. 
With the help of the outside sigmoid function, with the parameter $\lambda=10$,  $\mot_{GRW}$ is pushed away from $0.5$ towards $0$ and $1$. This modification is performed to obtain a continuous field, where the majority of the space is occupied by either low motility regions or high motility regions. One example is shown in Fig.~1.\textbf{(a) (main text)}. 
Furthermore, the final $\mot$ is  modified with appropriate sigmoid functions of the $y$-coordinate, to avoid low motility zones on the starting- or finish line. To test the versatility of the machine learning model, different methods for the generation of the motility field $\mot$ are employed~(see Sec.3B (main text)).

\subsection{Gray-Scott model for field generation}
\label{sec_app_grayscott}

The Gray-Scott model for reaction diffusion through integration of the Gray-Scott model for reaction-diffusion \cite{mcgough2004pattern,gray1990chemical} can be used to generate fields, that can be used as motility fields $\mot$. The equations model the time evolution of two scalar fields, that diffuse and interact with each other through the dynamics prescribed be the equations 

\begin{align}
    \frac{\partial u}{\partial t} = D_u \nabla^2 u + uv^2 + f(1-u),\\
    \frac{\partial v}{\partial t} = D_v \nabla^2 v + uv^2 - (f+k)v.
\end{align}

In Fig.~6~\textbf{(d)} (main text), we use a single solution, that fulfills $\muav \approx 0.557$, generated with the parameters
with $D_u=0.2$, $D_v=0.1$, $f = 0.0545$, $k = 0.06125$. $u$ and $v$ are initialised to $0.5$ and $0.25$ within a square of side length $L/5$ in the middle of the simulation box, respectively.
Random Gray-Scott fields, as used in Fig.~6~\textbf{(f)} (main text), can be generated for this choice of parameters, for instance, by initialising each node of the $u$-mesh to a random number in $[0,1]$ and the each node on the $v$-mesh to a random number in $[0,0.5]$.
The fields are discretised on a quadratic mesh of size $150\times150$. We solve the equations via an explicit forward Euler scheme in time, where we integrate for $10^4$ time-steps with $\Delta t = 1$, with periodic boundary conditions. Finally, after appropriate renormalisation, we use the resulting $v(\mathbf{r})$ as motility field $\mot$.

\section{Augmented Dijkstra's algorithm}
\label{sec_app_Dijkstras}

\begin{figure}[t]
\begin{center}
\includegraphics[width=0.5\linewidth]{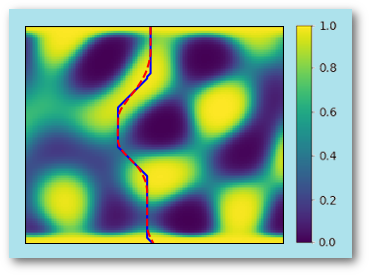}
\caption{Trajectories approximating the globally optimal trajectory obtained with \textit{Dijkstra's algorithm}. \textbf{Blue solid curve}: Trajectory obtained after employing \textit{Dijkstra's algorithm} on a discrete square grid in the simulation box, using local swimming time (Eq. (\ref{eq_app_t_swimming})) as grid distances. \textbf{Red dashed curve}: Approximation to the globally optimal trajectory, obtained after heuristic optimisation with a Monte-Carlo approach, using continuous trajectory nodes.}
\label{fig_app_Dijkstras}
\end{center}
\end{figure}


In order to obtain a visual reference for the ability of the swimmer to find a reasonable path through the environment, 
as well as having a quantitative benchmark for the difficulty of any environment $\mot$, we perform a global optimisation scheme with the help of \textit{Dijkstra's (shortest path) algorithm} \cite{dijkstra1959note}. E.W. Dijkstra initially conceptualised this algorithm for finding the shortest distance between two given nodes in a discrete network. A slightly more modern version, assumes an origin node and calculates the shortest distance from each node to the origin node.\\
In our version, we start with a network, where the positions of the nodes fall on the grid positions of the discretised motility field. We assign the node, closest to the designated starting position of the swimmer as origin node. The neighbours of the network are vertical, horizontal and diagonal next neighbours within the square grid. In accordance to the standard procedure of \textit{Dijkstra's algorithm}, we initialise the total distances of all nodes to the starting node to $\infty$, except itself. We update all total distances iteratively by evaluating the distances to all next neighbours, starting from the origin node, visiting each node once. The network distance $d_{i,j}$ between a pair of neighbouring grid nodes $i,j$ at $\mathbf{r}_i$,$\mathbf{r}_j$ we define by the Euclidian distance divided by the mean local swimmer velocity 

\begin{equation}
\label{eq_app_t_swimming}
    d_{i,j} = \frac{2\left\| \mathbf{r}_i - \mathbf{r}_j\right\|}{v_0(\mu(\mathbf{r}_i) + \mu(\mathbf{r}_j))}.
\end{equation}

Minimising with respect to the sum of the instantaneous swimming times $\sum_k d_{k,k+1}$, in a trajectory, yields an approximation to the solution of the functional, given in Eq. (5) (main text),
for any point in the simulation box $\mathbf{r}_1$ and the starting point $\mathbf{r}_0$ assuming the constraint, that the swimmer may only turn by angles that are multiples of $\pi/4$ . We choose as destination node, the node on the finish line, with the shortest distance to the origin node. Over the execution of \textit{Dijkstra's algorithm}, we save for each node the ancestor node, from which the shortest distance is obtained, such that we can backtrace optimal trajectories to the origin node in this step (see Fig.~\ref{fig_app_Dijkstras}, blue solid line).\\

In a second step, we augment the search for an optimal trajectory with a Monte-Carlo optimisation, to accommodate for the lack for continuous angles. More specifically, we perform $1000$ Monte-Carlo cycles in which we displace the individual nodes to optimise for the total travel time along the trajectory. We accept a trial displacement only, if it decreases the total travel time.
This heuristic supplementation results in smoother curves, and slightly reduces the total travel time (see. Fig.~\ref{fig_app_Dijkstras}, red dashed line). \\

In practice, we use this optimal first passage time as quantitative measure of the difficulty for the swimmer to maneuver through environments for a given set of parameters. More specifically, in the beginning of each training procedure, we use the current environment parameters, to generate a set of typical environments and solve it with the protocol described above. We multiply the obtained average first passage time with $100$ and use this as a termination criterion $T_{\mathrm{max}}$ for the individual episodes. We have found that this value of all the simulated environments typically corresponds to $ 10^{-3} \lesssim v_y /v_0 \lesssim 10^{-2}$.

\section{velocity state threshold {\large$\mu_0$}}
\label{sec_app_vel_state_thr}

\begin{figure}[t]
\begin{center}
\vspace{1cm}
\includegraphics[width=0.5\linewidth]{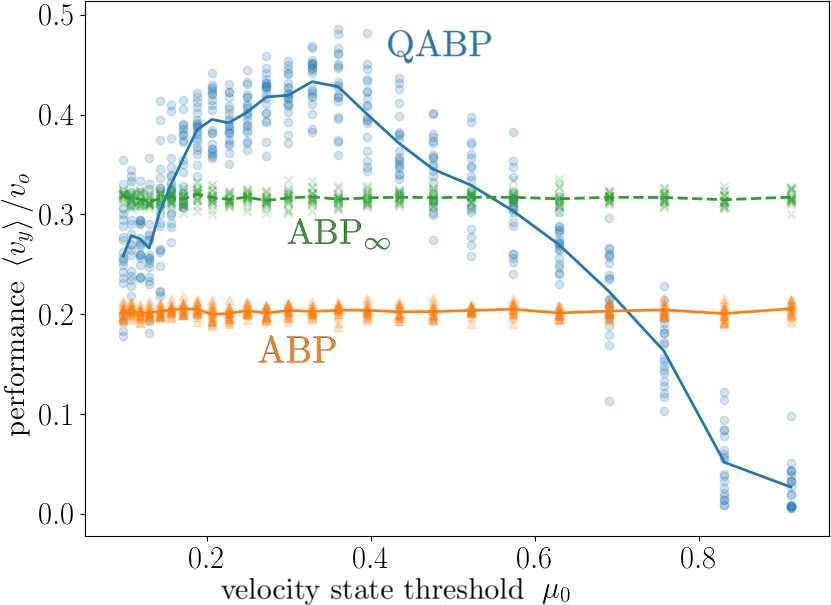}
\caption{
Performances $\perf$ of the QABP as a function of the velocity state threshold $\mu_0$ at $\mathcal{P}_{\mathrm{rot}} \approx 46$. The environments are generated with the help of as Gaussian random waves (see. Sec.~\ref{sec_env}) with $\muav \approx 0.557$. 
The scattered data data points indicate a simulation with $10^3$ trajectories each. The solid lines denote the averages over the respective ten data points per parameter set.
For reference, the figure shows the respective performances of ABP and ABP$_\infty$, which swims in a straight line.}
\label{fig_vel_state_threshold}
\end{center}
\end{figure}

In our reinforcement learning model, QABP receives information about its own instantaneous velocity $v_0 \mot$ in form of a binary discretisation. The state space of the agent encodes the information, whether the current $\mot$ is either above or below a threshold value $\mu_0$. As visible from the large peak just above $\mu_0$ in the velocity distribution of the QABP in Fig.~4 (main text), this threshold value directly determines the regions in the environment, that the swimmer aims to evade. Consequently, $\mu_0$ directly influences the performance $\perf$ of the swimmer (see Eq.~(6) (main text)). We show $\perf$ as a function of $\mu_0$ in Fig.~\ref{fig_vel_state_threshold} at $\mathcal{P}_{\mathrm{rot}} \approx 46$. The environments are generated with the help of Gaussian random waves (see. Sec.~\ref{sec_env}) with $\muav \approx 0.557$.
Even though independent of $\mu_0$, we show the performance of an active Brownian particle (ABP) and a particle swimming along a straight line trajectory (ABP$_\infty$) for reference.\\

As expected, the performance $\perf$ as a function of $\mu_0$ shows non-monotonic behaviour. For $\mu_0 \ll 1$, 
the lowest velocity, that the particle tolerates increases, hence $\perf$ increases. 
In the limit $\mu_0 = 1$, the QABP considers the whole environment as low motility to avoid and the algorithm ceases to function. As a result, 
for larger $\mu_0$, the relative differences between the instantaneous swimming velocities decrease in importance. Instead, the free space to maneuver dominates. In between the two competing limits, there is an optimal $\mu_0$ at approximately $0.3$ for this choice of environment and $\muav$. In the training of the QABP, we choose $\mu_0 = 0.25$.

\section{Reward for QABP}
\label{sec_reward_detail}
To impel a net upward motion of the swimmer, the last action is rewarded with $R_f = 100$ if the swimmer reaches the top of the box. To disincentivise the motion through low motility zones, an action is punished with $R_\mu = -0.5$, if it coincides with the termination of the episode due to entering a low motility region  $\mu < 0.125$. \\
Maneuvering with only local information in randomly generated fields almost inevitably leads to the encounter of dead end situations, depending on the difficulty of the environment, i.e. $\left < \mu \right >$.
In these cases, given only the two aforementioned rewards $R_f$, $R_\mu$, 
the swimmers 
trajectory coils up, and the swimmer is almost stuck in place. We suppress this issue, by 
introducing a third reward, which aims to incentivise the QABP towards trajectories with higher persistence. To this end, the swimmer memorises $100$ of its own local position, over a time of $100\tau_Q$. The swimmer measures its displacement with respect to the average of its saved positions.

In almost all trajectories, where the swimmer does not get stuck, the respective displacement lies around $1.0$. On the contrary, when it does get stuck, the displacement is close to $0$. Therefore we introduce an additional punishment.
If the displacement above falls below a value $0.5$, the current action is punished by $R_p = -0.4$. In general, the choice of this threshold value depends on the typical length scale of the environment. Here, employing a general threshold value sufficiently enhances the performance of the QABP across all simulated parameters.\\

\section{{\large$\epsilon$}-greedy policy}
\label{sec_eps_policy}
In every episode, the swimmer is given a random probability $\pran$, with which random actions are performed. Otherwise it uses $\mathcal{Q}$ as basis for decision. In episode $1$, $\pran$ is initialised as $1$ and linearly decreases until episode $10^3$, after which another $1000$ episodes are performed at $\pran = 0$, each in a new environment. At $\pran = 0$ the training procedure is considered to be finalised, $\mathcal{Q}$ to be converged, and the corresponding results are sampled. While sampling, the trajectories upon entering low motility zones are not terminated.
If the swimmer surpasses the upper bound $T_{\mathrm{max}}$, the trajectory is terminated and the first passage time is considered to be equal to $\infty$. This protocol of a decreasing $\pran$ takes into consideration the initial ignorance of the swimmer about his environment, and therefore the swimmer relies on random exploration. As the training proceeds, the decision matrix $\mathcal{Q}$ becomes gradually more reliable, and the swimmer targets the $\mathcal{Q}$ entries which are more likely to be the locally correct choices.

\section{Data storage}
The total amount of data storage that is needed for the decision matrix (Q-matrix) in our model necessitates $12 \times 12 \times 2 \times 3 = 864$ floating point numbers. The additional amount of data that is needed for the displacement reward $R_p$ incorporates $200$ floating point numbers for the hundred positions in two-dimensional space. The latter $200$ floating point numbers can be ommited in practical applications, by employing a finalised decision matrix $\mathcal{Q}$ obtained, e.g., in our computer simulation.

\bibliography{bibliog}